\documentclass[10pt,letterpaper,twocolumn]{article} 

\usepackage{ol}
\usepackage{hyperref}
\usepackage{amsmath}
\usepackage{epsfig}

\begin{document}

\twocolumn[

\title{Quantitative two-dimensional shadowgraphic set-up for high-sensitivity measurement of low-density laser-plasmas}

\author{Amrutha Gopal, Stefano Minardi, and Michael Tatarakis}

\address{Department of Electronics, Technological Educational Institute of Crete - Romanou, 3 -
GR73133 Chania, Greece}


\begin{abstract}We present a quantitative shadowgraphic method which can measure the density of a laser-generated plasma in air with sensitivity and resolution comparable or better than traditional interferometric techniques. Simultaneous comparison of both shadowgraphy and interferometry has been carried out allowing the experimental evaluation of the reliability of the shadowgraphic method. \end{abstract} 


]

Many branches of physics face the problem of the accurate characterization of small distributions ($d \simeq 20 \mu$m) of low density plasmas ($n_e < 10^{19}$ cm$^{-3}$), such as laser-produced plasma channels in dielectrics \cite{Dubietis06}, and coronal plasma jets generated in pulsed power devices (\textit{e.g.} X-pinch\cite{X-pinch}) or during high-intensity laser-solid interactions \cite{laser_plasma}.
Optical interferometry is the most favoured method for characterizing under-critical plasmas \cite{Matsumoto81,Borghesi96}. However, to probe a plasma slab of $20 \mu$m of thickness and density $n_e=10^{19}$ cm$^{-3}$ with visible light, a set-up with sensitivity far better than $\simeq\lambda/20$ is necessary, which is usually quite a demanding requirement. Moreover, many phase retrieval methods can not exploit the full resolution of the imaging system, the smallest resolvable detail is limited by the size of the interference fringes.
Recently, the plasma produced in laser-induced filaments in transparent solids has been characterized by means of measurements of the inverse bremsstrahlung absorption\cite{inverse_bremsstrahlung} of a probe pulse propagating perpendicularly to the filament\cite{Mao03,Sun05}.    
However, the density measurements obtained by these techniques heavily rely either on assumptions on the electron-ion collision time \cite{Mao03}, or on an independent interferometric measurement \cite{Sun05}, which is again limited in sensitivity and spatial resolution. 

In this Letter, we present a set-up based on quantitative refractive shadowgraphy \cite{Marcuse79a,Merzkirch87,deBruyn96} which can overcome these limitations and achieve a remarkable spatial resolution ($\simeq 25 \mu$m) and sensitivity ($\Delta n_e \simeq 5\cdot 10^{17}$ cm$^{-3}$), useful for the characterization of laser-induced plasmas in air.

According to geometrical optics in a homogeneous medium, optical rays belonging to a collimated unifrom beam of intensity $I_0$ propagate parallel to each other in the longitudinal direction $z$. The rays hitting a transparent, optically thin object are deflected by its transverse gradients of the refrective index and result in a transverse intensity pattern $I(x,y)$ (the shadowgram)at an obsevation plane placed at a distance $L$ from the object. 
By assuming negligible diffraction effects, small angle deflection, and the thickness of the sample $\Delta z\ll L$, it is possible to relate the variation of the transverse optical thickness of the sample to the shadowgram \cite{Merzkirch87,deBruyn96}:
\begin{equation}
\biggl( \frac{\partial^2}{\partial x^2} + \frac{\partial^2}{\partial y^2} \biggr)\int_0^{\Delta z} n(x,y,z) dz = -\frac{1}{L}\frac{I(x,y)-I_0}{I_0}
\end{equation}  
where $n(x,y,z)$ is the refractive index distribution of the phase object.
It is easy to recognize in Eq. 1 the Poisson equation: the unknown is the optical thickness of the object, while the source term is proportional to the normalized intensity variation of the shadowgram respect to the intensity of the incoming plane wave. 
For radially symmetric samples, the optical thickness distribution can be inverted by means of the Abel inversion transformation \cite{Abel} resulting in the actual radial distribution of refractive index $n(r)$, which in turn is proportional to the plasma density radial profile.
 
Theoretically, the accuracy of the shadowgraphic method is limited only by the impact of diffraction in free space propagation \cite{Trainoff02}. This can be minimized by taking $L$ small enough or by using a short-wavelength light source as object backlighter. From the practical viewpoint,  two main problems have to be solved in order to achieve a satisfactory accuracy, namely: \textit{i)} reduction of large-scale correlated noise sources (\textit{e.g.} air turbulence), and \textit{ii)} coding of fast numerical schemes for the resolution of the Poisson equation on large arrays of data.

To test the reliability of the shadowgraphic method, we used the set-up shown in Fig. 1.
A pulse from our amplifed sub-ns laser source (EKSPLA SL312, delivering 150 ps pulses at a wavelength of 1064 nm and energies up to 250 mJ) is expanded and divided into two channels by a semi-reflecting dielectric mirror. The reflected beam (pump) is focused in air using a $f=300$ mm lens to excite a line of plasma bubbles through laser breakdown. The transmitted beam (probe) is frequency doubled using a 1 cm thick KDP crystal and passed through an adjustable delay line (delay from 0 to 3 ns respect to the time of plasma bubble formation) and a spatial filter before probing orthogonally the plasma bubbles. The frequency conversion of the probe was required in order to reduce the impact of the inverse bresmstrahlung absorption. The probe is then collected by a $f=+50$ mm lens and divided in two channels by a 50/50 beamsplitter. On one channel, the shadowgram of a plane at distance $L=1.5$ mm from the bubbles is recorded. On the second channel a Nomarsky interferometer was built using 0.5 degree split angle Wollaston prism. The camera is focused on a plane containing the line of plasma bubbles to minimize distorsions \cite{Matsumoto81}. Both shadowgram and interferogram were recorded simultaneously using a dual frame grabbing software and by mutually synchronizing the cameras. Each measurement consisted of taking a frame of the plasma bubbles followed by a measurement of the background taken while stopping the pump beam. The shadowgraphic images were processed using a specifically developed fast Poisson solver using optimized successive over relaxation method, while the interferograms were processed by means of the IDEA software\cite{IDEA}. The whole set-up was enclosed in a box to avoid any disturbance in the air flow which could affect the shadowgraphic data.
\begin{figure}[htb]
\mbox{\epsfig{file=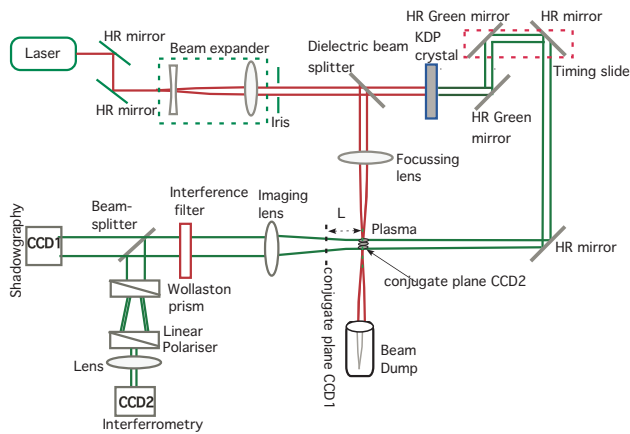, width=8.3cm}}
 \caption{Sketch of the experimental set-up}
\end{figure}
\begin{figure}[htb]
\mbox{\epsfig{file=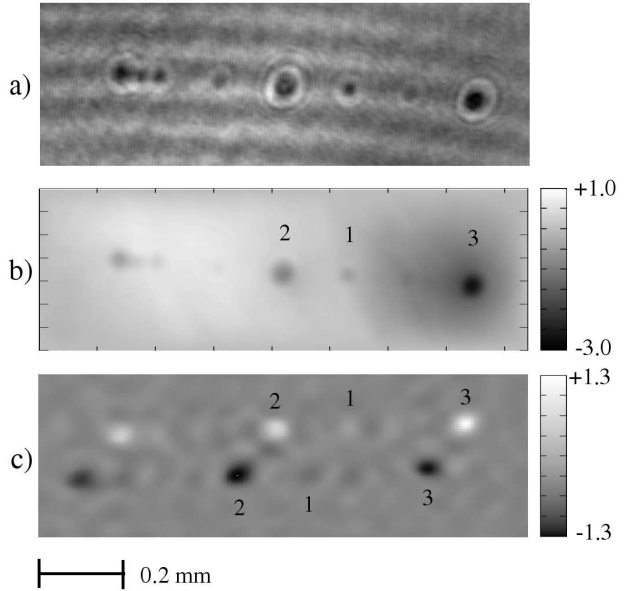, width=8.3cm}}
 \caption{a) Shadowgram of a chain of expanding plasma bubbles excited by a focused laser pulse  in air ($\tau_0 = 150$ ps, $\lambda = 1064$ nm, E = 10 mJ). Time delay of the probe pulse ($\tau_0 = 110$ ps, $\lambda = 532$ nm): +2.0 ns. Image taken at $L=1.5$ mm from the object plane. b) and c): Retrieved phase maps resulting from the shadowgraphic method and from the interferometric set-up, respectively. The double image in the interferometric map is a peculiarity of the modified Nomarski interferometer we used. The numbers refer to the bubbles displayed in Fig. 3.}
\end{figure}

A typical outcome of the processing of the data acquired by the two cameras is illustrated in Fig. 2. A qualitative comparison between the  interferometric and shadowgraphic measurements (Fig. 2.b and 2.c) reveals that the shadowgraphic method allows an accurate mapping of the array of laser-generated plasma bubbles in greater detail compared to the interferometric method.

Cross-section optical thickness profiles show that the shadowgraphic retrieval matches the interferometric measurement also on a quantitative basis (phase profiles, Fig. 3). 
For low to medium density plasma bubbles (Fig. 3.a and 3.b) the accuracy of the shadwographic retrieval is remarkably good. The peak values measured with both methods are the same within the measured large-scale fluctuation of the baselines ($\pm 0.1$ rad). Regarding the spatial width of the peaks, they match within the estimated resolution of the interferometric set-up (1 fringe separation = $28\mu$m). Abel inverted plasma density profiles (Fig. 3.d and 3.e) tend to enhance the differences near the peak value and reduce the impact of irregular baselines, however the agreement between the two methods is very good. The discrepancy found for low-density profiles has to be attributed mainly to the lower resolution of the interferometric set-up.    
For higher density profiles (Fig. 3.c), the shadowgraphic method apparently fails in reproducing accurately the interferometric phase data. In fact, the outcome of shadowgraphy shows the plasma peak sitting on a large-scale parabolic background affecting the absolute peak value of the phase shift. Notice however that the peak is indeed reproduced in great detail, as confirmed by the Abel transformed data (Fig. 3.f).

The appereance of a parabolic backgound in the shadowgraphic phase data can be explained quantitatively by assuming significant absorption of the probe beam in the plasma bubble due to inverse bremssthralung effect\cite{inverse_bremsstrahlung}. In fact, by suitable Fourier filtering of the interferograms we were able to reconstruct a low resolution $L=0$ shadowgram which clearly shows an absorption maximum of $12\pm 5$\% in the position of peak no.3. 
We simulated the impact of absorption by numerically generating shadowgrams of a Gaussian distribution of plasma in the frame of a paraxial beam propagation model. If the absorption losses are not taken into account in the background frame (\textit{i.e.} $I_0(x,y)=$const.) the retrieved phase map will develop a parabolic background such as the one observed experimentally. The best fit of the experimental data (not shown) is obtained for a peak absoprtion of 24\%. 
If the data are processed by using as a reference a shadowgram at $L=0$, the parabolic background can be removed almost completely (see Fig. 3.c).    

\begin{figure}[tb]
\centerline{\includegraphics[width=8.3cm]{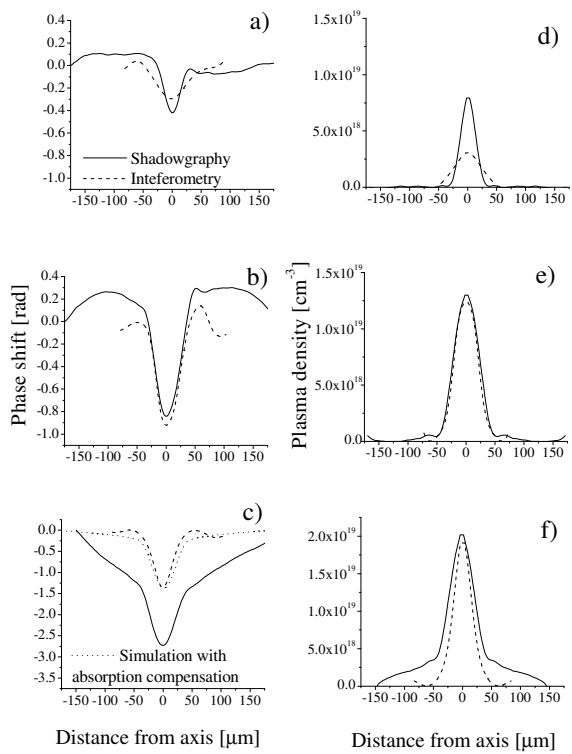}}
 \caption{Quantitative comparison of the shadowgraphic and interferometric method. a)-c) Phase profiles along the vertical direction of bubbles 1, 2 and 3 of Fig. 2, respectively. d)-f) Retrieved electron density profiles of the same bubbles.}
\end{figure}

The beam propagation model was also used to evaluate the ultimate sensitivity and resolution of the shadowgraphic method. The simulations\cite{ICOPS06} basically confirm previous results found in the literature\cite{Marcuse79a,Trainoff02} which predict a negligible impact of diffraction as long as \textit{i)} the observation distance $L$ is much smaller than the absolute value of the effective focal length of the phase object, and \textit{ii)} the far-field diffraction spot is much smaller than the transverse dimensions of the object. 
These requirements are in contrast to the requirement of large $L$ to achieve high sensitivity therefore, a tradeoff between resolution and sensitivity has to be found.  For a green light probe recorded with an 8-bit CCD camera, our simulations indicate that $L$ between 1.0 and 1.5 mm is a good compromise for the accurate retrieval of plasma bubbles with density in the range 0.5-20 10$^{18}$ cm$^{-3}$ and FWHM diameters greater than $\simeq$26-33 $\mu$m (accuracy better than 10\% in both peak phase and width retrieval).
The ultimate resolution of the method can be estimated by equating the spot size of the far-field diffraction to the actual size of the object. For $L$ in the chosen range, this gives a value of FWHM of the gaussian peak in between 20 and 26 $\mu$m.   

In conclusion, we have shown that quantitative refractive shadowgraphy is a reliable and useful technique for the accurate measurement of low-density, undercritical, radially symmetric plasma distributions. 
Compared to the existing literature, the novelty of our method resides on the possibility to retreive general two dimensional phase maps without the constraint of periodicity of the pattern\cite{deBruyn96}. We finally point out that, in contrast to shear interferometry, the method could in principle be scaled to probe pulses of a few optical cycles, thus opening interesting perspectives for the measurement of time resolved plasma dynamics during femtosecond pulse filamentation in dielectric media \cite{Dubietis06}.

The authors would like to aknowledge Nikos Skordilis for modifying the software of the frame-grabber and to thank Dr. Marco Potenza for stimulating discussions. This research work was supported by European Union through the Marie Curie Transfer of Knowledge grant DAIX Contract n. MTKD-CT-2004-014423.


\begin{thebibliography}{99}
\bibitem{Dubietis06} A. Dubietis, A. Couairon, E. Ku\v{c}inskas, G. Tamo\v{s}auskas, E. Gai\v{z}auskas, D. Faccio, P. Di Trapani \textit{Appl. Phys. B}  {\bf 84,} 439 (2006)
\bibitem{X-pinch} I.H. Mitchell, R. Aliaga-Rossel, R. Saavedra, H. Chuaqui, M. Favre, E.S. Wyndham {\bf 7,} 5140 (2000)
\bibitem{laser_plasma} M.S. Wei, F. Beg, E. Clark, A.E. Dangor, R.G. Evans, A. Gopal, K.W.D. Ledingham, P. McKenna, P.A. Norreys, M. Tatarakis, M. Zepf, K. Krushelnick \textit{Phys. Rev E} {\bf 70}, 056412 (2004)
\bibitem{Matsumoto81} Y. Matsumoto, K. Koyama, M. Tanimoto, M. Sugiura \textit{J. Appl. Phys.} {\bf 52,} 3259 (1981)
\bibitem{Borghesi96} M. Borghesi, A. Giulietti, D. Giulietti, L.A. Gizzi, A. Macchi, O. Willi \textit{Phys. Rev. E} {\bf 54,} 6769 (1996)
\bibitem{inverse_bremsstrahlung} S. Eliezer \textit{"The interaction of high-power lasers with plasmas"}, Chapter 5, Institute of Physics, 2002
\bibitem{Mao03} X. Mao, S.S. Mao, R.E. Russo \textit{Appl. Phys. Lett.} {\bf 82,} 697 (2003) 
\bibitem{Sun05} Q. Sun, H. Jiang, Y. Liu, Z. Wu, H. Yang, Q. Gong \textit{Opt. Lett.} {\bf 30,} 320 (2005) 
\bibitem{Marcuse79a} D. Marcuse \textit{Appl.Opt.} {\bf 18,} 9 (1979)
\bibitem{Merzkirch87} W. Merzkirch \textit{"Flow visualization"}, pag. 123--134, Academic Press, 1987
\bibitem{deBruyn96} R. de Bruyn, E. Bodenschatz, S.W. Morris, S.P. Trainoff, Y. Hu, D. S. Cannel, G. Ahlers \textit{Rev.Sci.Inst.} {\bf 67,} 2043 (1996)
\bibitem{Abel} I.H. Hutchinson \textit{"Principles of plasma diagnostics"}, Cambridge University Press, Cambridge 1987
\bibitem{Trainoff02} S.P. Trainoff, D.S. Cannell \textit{Phys. of Fluids} {\bf 14,} 1340 (2002)
\bibitem{IDEA} M. Hipp, J. Woisetschl\"ager, P. Reiterer, T. Neger \textit{Measurement} {\bf 36,} 53 (2004)
\bibitem{ICOPS06} S. Minardi, A. Gopal, K. Petridis, M. Tatarakis, \textit{"33th International Conference on Plasma Science"}, poster 2P-6, 4-8 June 2006, Traverse City (MI), USA 
\end{thebibliography}
\end{document}